\documentclass[sigconf]{acmart}

\usepackage{graphicx}
\usepackage{url}
\usepackage{tabularx}
\usepackage{caption}
\usepackage{subcaption}
\usepackage{lscape}
\usepackage{enumitem}
\usepackage{makecell}
\usepackage{url}
\usepackage{tabularx}
\usepackage{xcolor,colortbl}
\usepackage{multirow}
\usepackage{tikz}
\usepackage{comment}
\usepackage{xcolor}
\usepackage{tcolorbox}
\usepackage[title]{appendix}

\newcolumntype{Y}{>{\centering\arraybackslash}X}

\newcommand{\cready}[1]{\textcolor{black}{#1}}

\newlength{\RoundedBoxWidth}
\newsavebox{\GrayRoundedBox}
\newenvironment{GrayBox}[1][\dimexpr\linewidth-4.5ex]%
   {\setlength{\RoundedBoxWidth}{\dimexpr#1}
    \begin{lrbox}{\GrayRoundedBox}
       \begin{minipage}{\RoundedBoxWidth}}%
   {   \end{minipage}
    \end{lrbox}
    \begin{center}
    \begin{tikzpicture}%
       \draw node[draw=black,fill=black!10,rounded corners,%
             inner sep=2ex,text width=\RoundedBoxWidth]%
             {\usebox{\GrayRoundedBox}};
    \end{tikzpicture}
    \end{center}}

\AtBeginDocument{%
  \providecommand\BibTeX{{%
    \normalfont B\kern-0.5em{\scshape i\kern-0.25em b}\kern-0.8em\TeX}}}

\copyrightyear{2024}
\acmYear{2024}
\setcopyright{acmlicensed}
\acmConference[MSR '24]{21st International Conference on Mining Software Repositories}{April 15--16, 2024}{Lisbon, Portugal}
\acmBooktitle{21st International Conference on Mining Software Repositories (MSR '24), April 15--16, 2024, Lisbon, Portugal}
\acmDOI{10.1145/3643991.3644909}
\acmISBN{979-8-4007-0587-8/24/04}

\begin{document}


\title[What Can Self-Admitted Technical Debt Tell Us About Security? A Mixed-Methods Study]{What Can Self-Admitted Technical Debt\\ Tell Us About Security? A Mixed-Methods Study}




\author{Nicol\'{a}s E. D\'{i}az Ferreyra}
\affiliation{
  \institution{Hamburg University of Technology}
  \city{Hamburg}
  \country{Germany}
  }
\email{nicolas.diaz-ferreyra@tuhh.de}

\author{Mojtaba Shahin}
\affiliation{
  \institution{RMIT University}
  \city{Melbourne}
  \country{Australia}
  }
\email{mojtaba.shahin@rmit.edu.au}

\author{Mansooreh Zahedi}
\affiliation{
  \institution{The University of Melbourne}
  \city{Melbourne}
  \country{Australia}
  }
\email{mansooreh.zahedi@unimelb.edu.au}

\author{Sodiq Quadri}
\affiliation{
  \institution{Hamburg University of Technology}
  \city{Hamburg}
  \country{Germany}
  }
\email{sodiq.quadri@tuhh.de}

\author{Riccardo Scandariato}
\affiliation{
  \institution{Hamburg University of Technology}
  \city{Hamburg}
  \country{Germany}
  }
\email{riccardo.scandariato@tuhh.de}

\renewcommand{\shortauthors}{D\'{i}az Ferreyra et al.}

\begin{abstract}

Self-Admitted Technical Debt (SATD) encompasses a wide array of sub-optimal design and implementation choices reported in software artefacts (e.g., code comments and commit messages) by developers themselves. Such reports have been central to the study of software maintenance and evolution over the last decades. However, they can also be deemed as dreadful sources of information on potentially exploitable vulnerabilities and security flaws. \textbf{Objective}: This work investigates the security implications of SATD from a technical and developer-centred perspective. On the one hand, it analyses whether security pointers disclosed inside SATD sources can be used to characterise vulnerabilities in Open-Source Software (OSS) projects and repositories. On the other hand, it delves into developers' perspectives regarding the motivations behind this practice, its prevalence, and its potential negative consequences. \textbf{Method}: We followed a mixed-methods approach consisting of (i) the analysis of a preexisting dataset containing 8,812 SATD instances and (ii) an online survey with 222 OSS practitioners. \textbf{Results}: We gathered 201 SATD instances through the dataset analysis and mapped them to different Common Weakness Enumeration (CWE) identifiers. Overall, 25 different types of CWEs were spotted across commit messages, pull requests, code comments, and issue sections, from which 8 appear among MITRE's Top-25 most dangerous ones. The survey shows that software practitioners often place security pointers across SATD artefacts to promote a security culture among their peers and help them spot flaky code sections, among other motives. However, they also consider such a practice risky as it may facilitate vulnerability exploits. \textbf{Implications}: Our findings suggest that preserving the contextual integrity of security pointers disseminated across SATD artefacts is critical to safeguard both commercial and OSS solutions against zero-day attacks.

\end{abstract}



\maketitle

\section{Introduction} \label{sec:intro}

Software developers often face complex technical challenges that, for diverse reasons, they circumvent with spurious design workarounds and sub-optimal implementations \cite{guo2016exploring}. Such compromises are likely to result in a significant amount of Technical Debt (TD), namely low-quality artefacts that can impair the future maintenance and evolution of the system being developed. Different types of TD have been reported in the current literature, including code, design, requirements, and documentation. If not identified timely and properly managed, TD can incur extra work in the long-term, not to mention additional costs. Hence, it is deemed a critical issue for software development and considered ubiquitous to most software projects across all industry segments \cite{guo2016exploring}.


\subsubsection*{\textbf{Motivation}}
Security issues are closely related to TD as they are often grounded in quality compromises in the software being built \cite{rindell2019managing,siavvas2019empirical}. Simple coding mistakes and design flaws can lead to exploitable vulnerabilities that are time- and effort-intensive to repair later on \cite{camilo2015bugs,aldrich2023exploring}. Moreover, similar to TD, security incurs extra costs as it is frequently seen as an ``after-thought'' instead of an actual priority in the software development life cycle \cite{niazi2020maturity}. Therefore, recent studies have started looking at the intersection between TD and security with the aim of detecting and managing vulnerabilities in both code and the core architecture of information systems (e.g., \cite{siavvas2022technical, rindell2019managing, siavvas2019empirical, aldrich2023exploring}). That is, to leverage TD as a security indicator \cite{rindell2019managing} and for the automatic detection of code vulnerabilities \cite{siavvas2022technical}. Still, this research landscape is in its infancy and calls for further investigations at the interface of TD and software security.

A significant part of TD is explicitly reported by developers themselves in the form of ``self-admissions'', namely through software artefacts such as code comments \cite{li2023automatic}. Moreover, recent studies have regarded commit messages and pull requests as valuable sources of Self-Admitted Technical Debt (SATD) \cite{zampetti2021self}. These artefacts have been central for the analysis of different TD types beyond code, including requirement debt and documentation debt \cite{sierra2019survey}. However, SATD sources can also have major security implications when reporting flaws and vulnerabilities in the software. Particularly, an attacker could take advantage of \textit{security pointers} disclosed inside these sources (e.g., a code comment like \texttt{``TODO: Make it thread-safe''}) to spot exploitable weaknesses in the system threatening its availability, confidentiality and integrity. Nevertheless, and despite its importance, the security implications of SATD have not been investigated nor documented throughout the literature up to our knowledge.

\subsubsection*{\textbf{Contribution and Research Questions}}


This work elaborates on the implications of security pointers inside SATD instances across multiple sources. Particularly, it investigates whether explicit references to security vulnerabilities inside code comments, commits, issue reports, and pull requests can help characterise exploitable weaknesses inside Open-Source Software (OSS). We focused our study on OSS projects and repositories as they become easy targets of security attacks by making their artefacts visible and available to the public in general \cite{ladisa2023sok,lounici2021optimizing}. We conducted a mixed-methods analysis of data extracted from OSS repositories and a complementary online survey to address the following Research Questions~(RQs):




\begin{itemize}[leftmargin=*]
    \item \textit{\textbf{RQ1: Can SATD sources contain pointers to security weaknesses and vulnerabilities inside OSS repositories?}} To answer this RQ, we conducted an analysis on a preexisting dataset containing 8,812 SATD instances from multiple sources \cite{li2023automatic}. Using a keyword-based search, we extracted 546 security-relevant SATD candidates and inspected them manually for security pointers. As a result, we gathered 201 Security Self-Admitted Technical Debt (SSATD) observations that we considered for further analysis.
    \item \textit{\textbf{RQ2: Which particular weakness and vulnerabilities are frequently exposed through different SATD sources?}} For this RQ, we manually mapped each SSATD instance in our resulting dataset to Common Weakness Enumeration (CWE) identifiers. Overall, we identified 25 different CWE types across commit messages, code comments, pull requests, and issue reports. Moreover, 8 of these 25 CWEs appear listed among MITRE's Top-25 most dangerous software weaknesses in 2023.
    \item \textit{\textbf{RQ3: What is developers' perspective on security-related SATD in OSS repositories?}} We addressed this RQ through an online survey with OSS practitioners. We recruited a total of 222 participants and asked them about their tendencies to disclose security pointers inside SATD artefacts, their underlying motivations, and whether they considered this to be a risky practice. Our results show that many practitioners engage with this practice, although they recognise it as potentially risky. In total, we identified nine motives that drive developers to disclose security pointers inside SATD artefacts (e.g., ‘promote a security culture’) and three potential risks stemming from them (e.g., ‘security misconceptions’).

\end{itemize}

The remainder of this paper is organised as follows. Section \ref{sec:background} provides the background and summarises the related work. We explain our research methodology in Section \ref{sec:methodology}, followed by reporting the findings in Section \ref{sec:results}. Section \ref{sec:discussion} reflects on the findings and provides implications for research and practice. We discuss the possible threats of our study and adopted mitigation strategies in Section \ref{sec:validity}. Section \ref{sec:conclusion} concludes the paper and discusses some future research directions.

\section{Background and Related Work}\label{sec:background}

 
\subsubsection*{\textbf{Security and Technical Debt}} As mentioned, only a few publications in the current literature explore the interface between software security and TD. \citet{rindell2019sec} conducted a seminal work addressing the role and importance of TD for managing emerging security risks in software projects. They propose extending traditional TD management frameworks with (i) means for assessing the risks of TD items and (ii) security-based tools for unveiling unintentional TD instances (i.e., those caused by unawareness or sheer recklessness). Overall, they suggest that tools/methods such as code reviews and threat modelling are suitable to spot internal quality issues in software requirements, coding, architecture, and testing \cite{rindell2019managing}. Alongside, \citet{martinez2021secd} conducted a literature review to understand the main sociotechnical characteristics of ``security debt'' and its impact on the software development life cycle. They acknowledge that TD can entail security implications at different levels (e.g., code and architecture) and, given its potential damage, should be highly prioritised. 

Prior work has also elaborated on the interplay between TD and security flaws, namely on their role for the (automatic) identification of exploitable weaknesses. For instance, \citet{izurieta2018td} proposed a method to assess and prioritise TD instances linked to security flaws using the Common Weakness Enumeration (CWE) and its scoring mechanism. This approach was further investigated and extended in \cite{izurieta2019td} by mapping CWEs to malicious attack tactics. Similarly, \citet{siavvas2019empirical} explored TD as a software security indicator and used it later on as a feature for predicting vulnerabilities at both project and class levels \cite{siavvas2022technical}. They conclude that TD pointers can help improve the performance of Machine Learning (ML) models for vulnerability prediction. Recent work by \citet{aldrich2023exploring} has investigated security-related TD in Question-Answer (Q\&A) platforms like Stack Overflow (SO). Their findings show that around 38\% of security questions in SO are related to TD, being \textit{encryption} along with \textit{cryptography} among the most popular discussion topics.


\subsubsection*{\textbf{Security and Self-Admissions}} Research on self-admissions, on the other hand, seeks to draw a more complete picture of the TD landscape by examining additional sources of actionable information. It is worth mentioning that, despite that SATD literature is sparse, most of it has focused on the analysis of TD through code comments. For example, \citet{maldonado2015satd} investigated the different types of TD (e.g., design or documentation debt) developers disclose inside their code comments and provided a set of heuristics for their identification. Likewise, \citet{huang2018identifying} and \citet{de2020identifying} applied text mining techniques for the automatic extraction of SATD inside comments. \citet{russo2022msr} conducted one of the few investigations addressing security in SATD and observed that more than 55\% of self-admissions in the form of code comments correspond to insecure implementations/snippets. 

Recent work has stressed the importance of scrutinising alternative SATD sources such as commit messages and pull requests to better characterise TD in software projects \cite{zampetti2021self, li2023automatic}. Under this premise, \citet{li2023automatic} ran a large-scale study on SATD addressing different sources, including code comments, commit messages, pull requests, and issue reports across 103 open-source projects. Their results show that SATD is evenly spread among all these sources with a high prevalence of code/design debt. Still, up to the extent of our knowledge, multi-source studies at the interface between SATD and security are missing in the current literature. Particularly, the impact of explicit \textit{security pointers} across multiple SATD instances (e.g., comments or commit messages suggesting the presence of vulnerabilities) has not been actively explored and calls for further investigations.

\subsubsection*{\textbf{Documentation Practices}} 
Research has also addressed developers' documentation and repayment practices of SATD. For instance, \citet{shmerlin2015document} studied the drivers behind code documentation and observed that \textit{comprehension}, \textit{order} and \textit{quality} are among developers' primary motivations. Conversely, they observed that the perceived \textit{time} and \textit{effort} required to document code often act as deterrents for SATD. Code reviews were also shown to play a key role in the emergence of SATD. Particularly, a study by \citet{kashiwa2022empirical} revealed that around 28-48\% of self-admissions inside comments extracted from Qt and Open Stack systems were introduced during reviewing activities. Alongside, a survey by \citet{xavier2022documentation} investigated the motivations that drive developers to (i) introduce and (ii) repay SATD reported in issues. Results show that \textit{timely delivery} is often a cause for the former, whereas \textit{reducing the costs} and interests associated with TD may influence the latter.


Although the reasons why developers take and later repay TD are well documented in the current literature, little is known about their motivations to report it in the form of self-admissions. That is, which are the underlying factors that either encourage or discourage developers from reporting technical shortcuts across different software artefacts. An exploratory study on SATD in R packages by \citet{vidoni2021self} provides some insights, suggesting that self-admissions are often introduced as ``self-reminders'', ``warnings'', or ``to facilitate future planning''. Still, other intrinsic and extrinsic motivations, such as perceived rewards, risks, and social norms, remain unexplored and demand a thorough investigation at both software engineering and security levels.


\section{Methodology}\label{sec:methodology}

\begin{figure*}[ht]
\includegraphics[width=0.75\textwidth]{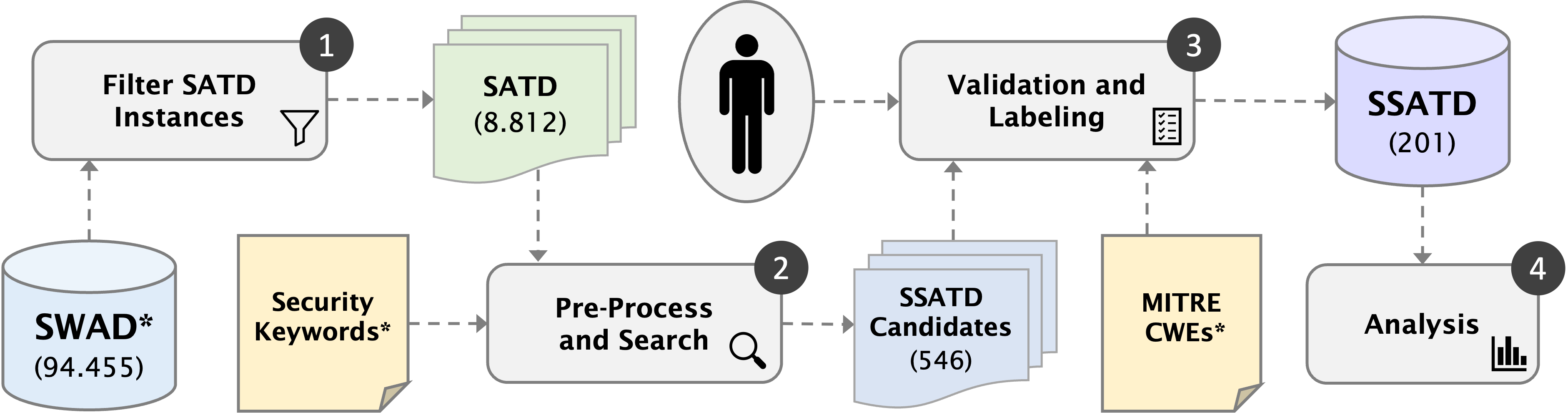}
\caption{Methodology applied for curating the SSATD dataset. \underline{Note}: External input is indicated with an asterisk.}
\label{fig:method}
\end{figure*}

To answer the RQs presented in Section~\ref{sec:intro}, we followed a mixed-methods approach encompassing (i) the analysis of OSS repositories and (ii) an online survey with practitioners. The following sections describe the steps we followed at each stage along with the instruments employed during the different data acquisition and processing activities.


\subsection{Repositories Study} \label{sec:repo_study}

This study consisted of identifying and analysing security pointers disclosed inside SATD artefacts. For this, we used a preexisting SATD dataset as a reference and conducted a keyword-based search enhanced with a series of manual validations and annotations.

\subsubsection{Reference Dataset} 
In a recent publication, \citet{li2023automatic} presented the results of a large-scale study on SATD in OSS projects and released a dataset for the automatic identification of self-admissions. The dataset contains \textbf{94,455 software artefacts} from which 5,000 correspond to commit messages and another 5,000 to pull requests collected from 103 Apache OSS projects. It also includes 23,180 issue sections collected during a prior study \cite{li2022identifying} and 62,275 code comments from a dataset curated by \citet{da2017using}. All instances in the final dataset are labelled either as SATD or non-SATD, making it suitable for the purpose of our research.

\begin{table}[t]
   \caption{Number and type of SATD instances in the \textit{reference}, \textit{candidates}, and \textit{final} datasets.} 
    \centering
    \begin{tabularx}{\linewidth}{c c c c}
    \toprule
     \textbf{Source} & \textbf{SATD} & \textbf{SSATD Candidates} & \textbf{Final SSATD}\\ \midrule
      Comments   & 4,071 & 180 & 29\\ 
      Commits   & 747 & 45 & 33\\
      Issues   & 3,276 & 284 & 123\\
      Pull requests   & 718 & 37 & 16\\ \midrule
      \textbf{Total}   & 8,812 & 546 & 201 \\ \bottomrule
    \end{tabularx}
    
    \label{table:dataset}
\end{table}

\subsubsection{Generation of SSATD Candidates} \label{sec:mthd_cand}
Figure~\ref{fig:method} illustrates the steps we followed for curating a dataset of Security Self-Admitted Technical Debt (SSATD) instances. First, we took the \textit{software artefacts} dataset (SWAD) of \citet{li2023automatic} and filtered the SATD instances by checking the corresponding label value (i.e., SATD or non-SATD). As shown in Table~\ref{table:dataset}, we gathered \textbf{8,812 self-admissions} from which 4,071 correspond to code comments, 747 to commit messages, 3,276 to issue sections, and 718 to pull requests. Next, we conducted a keyword-based search to detect SSATD candidates using a list of security terms elaborated by \citet{croft2022empirical}. This list is an updated version of the one proposed by \citet{le2020puminer} and contains 288 security keywords (e.g., `leak', `thread-safe'), making it one of the most extensive ones documented in the literature. Although this type of filtering has known limitations (e.g., it can produce false positives), we considered it appropriate at this stage given the size of our dataset and the relatively short text length of the self-admissions. We applied the keyword search after pre-processing each SATD instance (i.e., by removing all punctuation characters) and obtained a total of \textbf{546 SSATD candidates}.


\subsubsection{Manual Validation and Labelling}\label{sec:mthd_val} Two of the authors manually validated the dataset of SSATD candidates by checking for false positives. Both of them had prior experience in software security (one slightly senior to the other), so they conducted this task based on their own knowledge and technical expertise. As a general rule, they had to determine whether the information disclosed in the artefact they assessed could indicate the presence of a security flaw. Each author applied this criterion independently and labelled each of the 546 SSTD candidates as a \textit{true positive} (TP) or \textit{false positive} (FP), accordingly. We measured the level of agreement between the classifications of both authors using Cohen's Kappa. The values obtained indicate a `substantial' agreement with a coefficient of +0.65 \cite{landis1977measurement}. The authors then discussed, negotiated, and solved the discrepancies in their annotations on a one-by-one basis during a joint session. This led to a final dataset of \textbf{201 SSATD instances}, which they mapped jointly to Common Weakness Enumeration (CWE) identifiers in a follow-up session. Supported by MITRE, the CWE initiative consists of a public, community-developed catalogue with information about more than 900 types of software and hardware weaknesses. The mapping was conducted following an opportunistic approach in which the authors browsed the CWE database\footnote{\url{https://cwe.mitre.org/}} using the keywords found on each SSATD item as a guide. Overall, we identified \textbf{26 different CWEs} across the items included in our final dataset.

\subsection{Survey Study} \label{sec:survey_study}

To gather insights into developers perceptions about SSATD we conducted an online survey addressing different behavioural and motivational aspects. It consisted of 7 closed-ended and 2 open-ended questions that took participants around 8 minutes to respond. 


\subsubsection{Structure}\label{sec:surv_str} We designed our survey based on the findings collected during the repositories study. Particularly, we used information about the most frequent CWEs in the SSATD dataset to elaborate a set of representative examples that guided participants throughout the study. \cready{These were CWE 119 for commit messages, CWE 20 for code comments, and CWE 362 for both issue sections and pull requests}. Such examples (e.g., the excerpt of an issue mentioning ``\texttt{cache producer is not thread safe}'') aimed to prime participants about situations they may have encountered while developing software. That is, about cases in which security pointers get disseminated across code comments, commit messages, pull requests, or issue sections. We referred participants to these examples on each of the reminder survey questions and asked them to answer based on their own experience working on OSS projects:
\begin{itemize}[leftmargin=*]
    \item \textbf{SSATD Prevalence:} To understand SSATD prevalence in OSS repositories, we asked developers (i) how often do \textit{they (themselves)} disclose security pointers inside comments, commits, issues, or pull requests (\textbf{Q1}), and (ii) how often do they see \textit{other developers} doing so (\textbf{Q2}). These two questions were ranked using a five-point scale as ``Never'', ``Rarely'', ``Sometimes'', ``Often'', and ``Always''.
    \item \textbf{SSATD Motivations:} We also elicited participants' motivations to engage in this practice by asking them to assess a set of statements. \citet{assal2019think} suggest that developers' (i) work environment, (ii) sense of awareness, (iii) perceived rewards, and (iv) perceived negative consequences can drive them to engage in security-related practices. Hence, we elaborated one statement (\textbf{Q3}-\textbf{Q6}) for each of these 4 prospective motivations (e.g., \textit{``I introduce security pointers inside commits, issues, comments, or pull requests because this is a standard practice at work''}) and asked participants to rate them using a six-point scale (``Strongly Agree'', ``Agree'', ``Somehow Agree'', ``Somehow Disagree'', ``Disagree'', and ``Strongly Disagree''). We also added an open-ended question (\textbf{Q7}) to ask them whether they could think of any other drivers besides the ones mentioned in the statements.
    \item \textbf{SSATD Risks:} Finally, we asked participants about their perceptions of risks stemming from the disclosure of security pointers inside software artefacts. Particularly, we asked them to indicate how much they agreed (or disagreed) with the following statement: \textit{``I think security pointers inside commits, issues, comments, or pull requests may introduce risks in the software being deployed''} (\textbf{Q8}) using a six-point scale (``Strongly Agree'', ``Agree'', ``Somehow Agree'', ``Somehow Disagree'', ``Disagree'', and ``Strongly Disagree''). As with the motivation statements, we also included an open-ended question (\textbf{Q9}) to ask them which specific risks they think such pointers may introduce in OSS projects.
\end{itemize}

A subset of the main survey questions can be found in \autoref{appendix}. The complete survey instrument is available in the paper's \hyperref[sec:replication]{\textbf{Replication Package}}.

\subsubsection{Population and Recruitment} \label{sec:recruitment}

Participants were recruited via Prolific\footnote{\url{https://www.prolific.com/}}, a crowdsourcing platform frequently adopted to conduct empirical studies in software engineering \cite{tahaei2022recruiting,kaur2022recruit}, and prescreened using its built-in qualification features. We targeted individuals who reported to have (i) knowledge of software development techniques and (ii) computer programming skills. We validated these qualifications through a short technical questionnaire similar to the one proposed by \cite{krause2023pushed}. It consisted of 3 closed-ended questions, the first one asking whether the participants had contributed to OSS projects in the past (answer options: ``yes''/``no''), while the remaining two asked them to indicate the \textit{parameter} and \textit{output} of a given pseudocode, respectively (answer options: multiple choices). This screening questionnaire took around 2 minutes, for which we rewarded participants with 0.2 GBP each. Only those who passed it were invited to take the \textit{main survey} (Section~\ref{sec:surv_str}) and received an additional compensation of 1.10 GBP. As a general rule, participants had to be at least 18 years old to join both parts of the study (i.e., screening questionnaire and main survey), have taken part in at least 10 other studies in Prolific, and have a minimum approval rate of 98\%. Two attention questions were also included in the main survey to identify and discard answers from unengaged participants.

After running a pilot study with 10 participants, we observed that around 30\% passed the technical questionnaire. Based on this ratio and our study budget, we targeted a sample of size N=200. Hence, we repeated the process with 740 additional participants, from which 242 were deemed suitable for undertaking the main survey (i.e., after validating their technical knowledge and prior experience with OSS projects). \cready{Of these 242 participants, 15 did not accept our invitation to take the main survey, resulting in 227 main survey submissions. After checking for completeness and attention questions, a total of \textbf{222 answers} remained valid for analysis}.


\subsubsection{Ethical Considerations}

We received approval from the Ethics Committee of the German Association for Experimental Economic Research to conduct this study and followed the guidelines prescribed in the Declaration of Helsinki. We informed participants about the study procedure along with details concerning the privacy of their personal data before joining the experiment. We also asked for their informed consent and allowed them to withdraw at any time without their answers being recorded.


\section{Results}\label{sec:results}

We conducted multiple analyses of the data collected through the repositories and survey studies. Our main findings and characteristics of the resulting datasets are summarised in the following subsections. 

\subsection{Security Pointers in SATD (RQ1 and RQ2)}\label{sec:res_pointers}

As mentioned in Section~\ref{sec:mthd_val}, we gathered 201 SSATD instances, which represent 2.28\% of the total SATD available in the reference dataset. Since instances included in the reference dataset were labelled with the type of TD they refer to, so are the SSATD cases included in ours. Table~\ref{table:satd_types} shows that most of them correspond to sub-optimal choices taken at the code/design levels (72.1\%), followed by unsatisfied or partially-implemented requirements (16.9\%). Uncorrected known defects and poor testing come next with 4\% of the total cases each. Other TD types like documentation, architecture and build debt represent another 3\% altogether. We observe that most of the code/design and requirement debt is concentrated around issue sections, whereas defects are often self-admitted through code comments. On the other hand, commit messages seem to gather most of the test debt cases and half of the documentation debt ones. The rest of the TD types, namely architecture and build debt, were found inside one issue section and one commit message, respectively.


\begin{table}[!h]
    \caption{Types of TD found in the SSATD dataset.}
    \centering
    \begin{tabularx}{\linewidth}{@{}c c c c c | Y@{}}
    \toprule
     \textbf{SATD Type} & \textbf{PR} & \textbf{CC} & \textbf{IS} & \textbf{CM} & \textbf{TOTAL}\\ \midrule
      Defect Debt   & 0 & 7 & 1 & 0 & 8\\ 
      Code/Design Debt   & 12 & 17 & 96 & 20 & 145\\ 
      Requirement Debt   & 3 & 5 & 21 & 5 & 34\\
      Test Debt  & 1 & 0 & 2 & 5 & 8\\
      Documentation Debt  & 0 & 0 & 2 & 2 & 4\\ 
      Architecture Debt  & 0 & 0 & 1 & 0 & 1\\
      Build Debt & 0 & 0 & 0 & 1 & 1\\ \bottomrule
      \multicolumn{6}{l}{\parbox{\linewidth}{\vspace{0.8ex}\footnotesize \underline{Note}: PR=Pull Request; CC=Code Comment; IS=Issue Section; CM=Commit Message}}
    \end{tabularx}

    \label{table:satd_types}
\end{table}

\begin{figure}[!t]
\includegraphics[width=\linewidth]{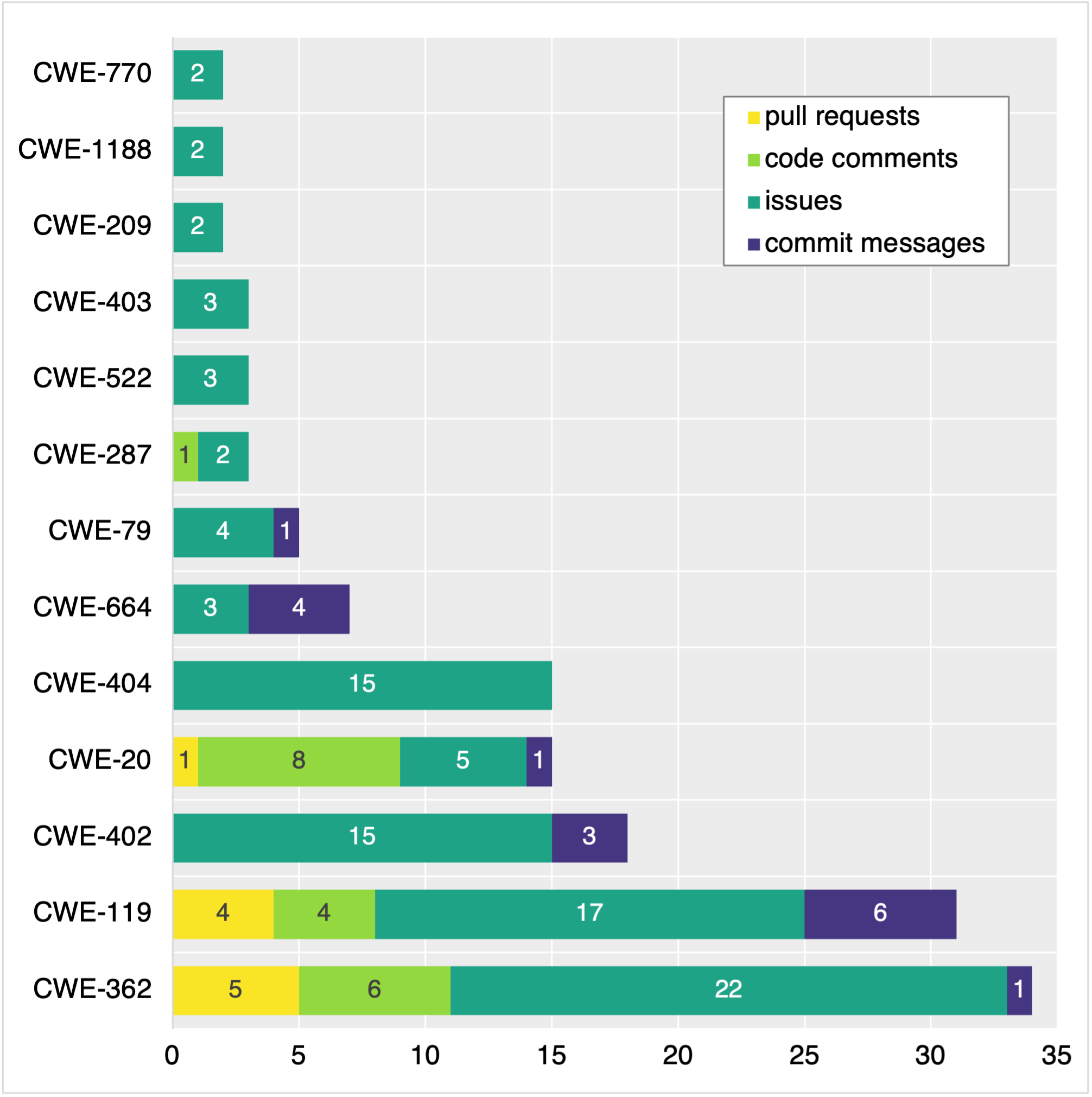}
\caption{Distributions of CWE across SSATD artefacts. \underline{Note}: CWEs with a frequency < 2 are not displayed in the graph.}
\label{fig:cwe_freq}
\end{figure}

Table~\ref{table:ssatd_examples} summarises the frequency of each CWE we identified in our final dataset, along with a representative example. Overall, we spotted 25 different CWE pointer types, from which 8 appear in MITRE's Top-25 list of the most dangerous software weaknesses in 2023 (i.e., CWEs 362, 119, 20, 79, 287, 89, 352, and 798). We observe that \textit{CWE-362: `Concurrent Execution using Shared Resource with Improper Synchronization (Race Condition)'} is the most frequent one, followed by \textit{CWE-119: `Improper Restriction of Operations within the Bounds of Memory Buffer'} with 34 and 31 cases, respectively. As shown in Figure~\ref{fig:cwe_freq}, these two most popular security pointers are often disclosed inside issue sections. However, we also found them inside pull requests, code comments, and commit messages with clear references to (i) non-thread-safe implementations in the case of CWE-362 and (ii) memory leakages in the code for CWE-119.

\begin{table*}[t]
\fontsize{8}{8}\selectfont
\setlength{\tabcolsep}{5pt}
\def\arraystretch{1.5}
\caption{Examples and frequencies of security pointers included in the final SSATD dataset.}
    \centering
    \begin{tabularx}{\linewidth}{p{6.5cm} | c | p{9.58cm}}
    \toprule
     \textbf{Pointer to} & \textbf{Freq.} & \textbf{Example} \\ \midrule
    \rowcolor{gray!25}\textbf{CWE-362:} Concurrent Execution using Shared Resource with Improper Synchronization ('Race Condition')  & 34 &  ``\texttt{Cache producer is not thread safe}'' [IS] \\ 
    
    \textbf{CWE-119:} Improper Restriction of Operations within the Bounds of a Memory Buffer & 31 &  ``\texttt{Memory leak from the main thread on shutdown}'' [CM]\\ 
    
    \rowcolor{gray!25}\textbf{CWE-402:} Transmission of Private Resources into a New Sphere ('Resource Leak') & 18 &  ``\texttt{Fix Python TSocket resource leak on connection failure}'' [IS]\\ 
    
    \textbf{CWE-20:} Improper Input Validation & 15 &  ``\texttt{TODO: Validate port as numeric}'' [CC]\\ 
    
    \rowcolor{gray!25}\textbf{CWE-404:} Improper Resource Shutdown or Release & 15 &  ``\texttt{Potential resource leak in due to unclosed stream}'' [CC]\\ 
    
    \textbf{CWE-664:} Improper Control of a Resource Through its Lifetime & 7 &  ``\texttt{Fix open file leak in log cleaner integration tests}'' [CM]\\ 
    
    \rowcolor{gray!25}\textbf{CWE-79:}	Improper Neutralization of Input During Web Page Generation ('Cross-site Scripting') & 5 &  ``\texttt{The ``)]\}'' prefix in the JSON response is intended ... It's a security measure to prevent XSSI}'' [IS]\\ 
    
    \textbf{CWE-287:} Improper Authentication & 3 & ``\texttt{TODO: Add dialog to inform user that the smtp server does not support authentication}'' [CC]\\ 
    
    \rowcolor{gray!25}\textbf{CWE-522:} Insufficiently Protected Credentials & 3 &  ``\texttt{Plain-text information leak of due to autosuggest}'' [IS]\\
    
    \textbf{CWE-403:} Exposure of File Descriptor to Unintended Control Sphere ('File Descriptor Leak') & 3 & ``\texttt{Metrics system FileSink can leak file descriptor}'' [IS] \\ 
    
    \rowcolor{gray!25}\textbf{CWE-209:} Generation of Error Message Containing Sensitive Information & 2 & ``\texttt{A new user was trying to perform a ``repo init'' for the first time on his workstation and saw the following output: ... ``fatal: not a Gerrit project'' ...I later discovered that the user was not a member of previously established LDAP groups with read permissions in Gerrit. That being the case, I would have expected to see a "permission denied" type message, not a message stating that the requested project did not exist, when in fact it did...}'' [IS]\\ 
    
    \textbf{CWE-1188:} Insecure Default Initialization of Resource & 2 &  ``\texttt{SSHD default MAC/KEX/cipher settings are rather insecure}'' [IS] \\ 
    
    \rowcolor{gray!25}\textbf{CWE-770:} Allocation of Resources Without Limits or Throttling & 2 & ``\texttt{The JobTracker can be prone to a denial-of-service attack if a user submits a job that has a very large number of tasks ... It would be nice to have a configuration setting that limits the maximum tasks that a single job can have}'' [IS] \\ 
    
    \textbf{Other CWEs with just one observation} (CWE: 89, 121, 668, 319, 312, 352, 477, 326, 798, 295, 172, 825) & 12 & ``\texttt{silly hack to avoid stack overflow}'' [CC | CWE-121]\\ 
    
    \rowcolor{gray!25}\textbf{Generic security pointers} & 69 & ``\texttt{The original logic looks not safe. Do you think if there is any other better way?''} [PR]\\ \bottomrule
    
    \multicolumn{3}{l}{\underline{Note}: PR=Pull Request; CC=Code Comment; IS=Issue Section; CM=Commit Message}
    \end{tabularx}
    
    \label{table:ssatd_examples}
\end{table*}

We also encountered pointers to \textit{CWE-402: `Transmission of Private Resources into a New Sphere (Resource Leak)'} and \textit{CWE-20: `Improper Input Validation'} with a relatively high frequency of 18 and 15 cases, respectively. CWE-20 was found in all artefact types but most frequently inside issue sections and code comments. These pointers often refer to missing or improper input validation, as in a code comment mentioning ``\texttt{TODO: Validate port as numeric}''. On the other hand, the majority of CWE-402 pointers were grouped around issue sections, with some of them also disclosed in the form of commit messages. Such is the case of an issue section in which a developer asks to fix a resource leak caused by a connection failure (see example in Table~\ref{table:ssatd_examples}). We also found 15 pointers to \textit{CWE 404: `Improper Resource Shutdown or Release'}, all of them corresponding to commit messages. In this case, pointers describe situations in which resources are not properly released and end up exposing sensitive data in subsequent memory allocations.

\textit{CWE-664: `Improper Control of a Resource Though its Lifetime'} and \textit{CWE-79: `Improper Neutralisation of Input During Web Page Generation (Cross-site Scripting)'} were also found inside commit messages and issue sections but with a lower frequency. Particularly, we retrieved 7 pointers to CWE-664 and 5 to CWE-79. In the case of the former, the pointers describe cases where resources are not properly created, used, or destroyed, which can lead to exploitable system states and unexpected behaviours. For the latter, the issue sections or commit messages make reference to the presence of Cross-site Scripting (XSS) vulnerabilities due to improper neutralisation of user-controllable input. The remaining CWE pointers cover 18 different CWEs and have 3 or less observations each (e.g., CWE-209 and CWE-121). They account for 27 cases in total, most of them found inside issue sections. We also found generic pointers whose content is not specific enough to map them to concrete CWEs. For instance, a pull request describing ``\texttt{The original logic looks not safe. Do you think if there is any other better way?''} suggests that the current implementation is not entirely secure but does not offer much detail about the specific threats or flaws in the code. All in all, such generic pointers add up to 31.22\% of the observations in the SSATD dataset.

\subsection{Developers' Perspective (RQ3)} \label{sec:survey_results}

Of 222 study participants, 190 were male, 29 were female, and 3 were non-binary. Around 23.9\% reported having more than 5 years of experience working with OSS projects, 13.9\% less than a year, and the rest between 1 and 5 years. Regarding their experience with closed-source development, about 31.1\% had more than 5 years of working experience, 23.9\% less than 1, whereas the remainder reported having between 1 and 5 years. A complete overview of the sample's demographics is shown in Table~\ref{tab:demographics}.

\subsubsection{SSATD Prevalence} 
\begin{figure}[!h]
\includegraphics[width=\linewidth]{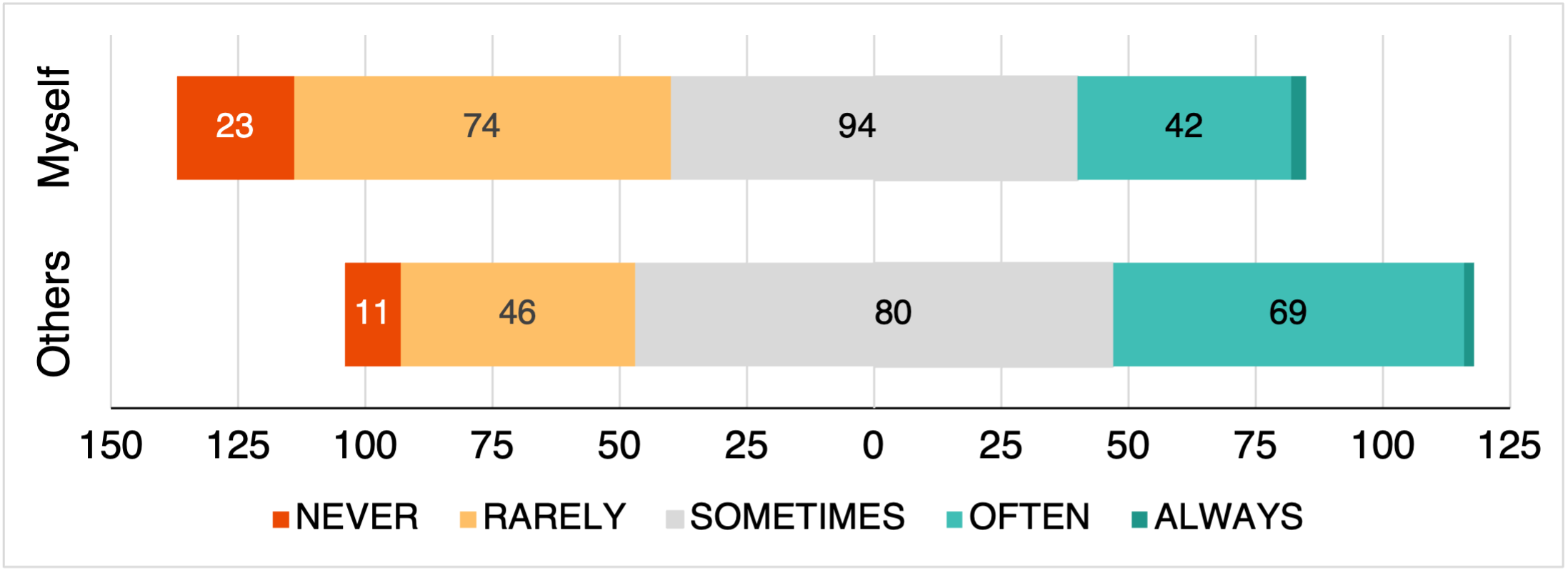}
\caption{Prevalence of security pointers in SATD. \underline{Note}: Frequencies < 10 are not labelled in the graph.}
\label{fig:prevalence}
\end{figure}

We can observe from Figure~\ref{fig:prevalence} that the majority of participants have disclosed security pointers (themselves) inside SATD artefacts at some point (89.6\%). That is, either \textit{rarely}, \textit{sometimes}, \textit{often}, or \textit{always}. Still, a relatively high percentage reported to \textit{rarely} engage in this practice (33.3\%) or not having engaged on it at all while working with OSS repositories (10.4\%). We also observe that most respondents have seen at some point another developer adding pointers inside commits, code comments, issue sections, or pull requests (95\%). However, 20.7\% of participants claimed to have \textit{rarely} identified such behaviour in others, whereas just 5\% reported not to have seen it at all.

\begin{table}[t]
\caption{Survey self-reported demographic data.}
\label{tab:demographics}
\centering
\begin{tabularx}{\linewidth}{lXcc}
\toprule
\textbf{Demographic} & \textbf{Ranges} & \textbf{Freq.} &\textbf{\%} \\
\midrule
\multirow{3}{*}{Gender}
& Male          & 190    & 85.6\%   \\
& Female        & 29    & 13.1\%   \\
& Non-Binary    & 3     & 1.3\%   \\          
\midrule
\multirow{4}{*}{\makecell[l]{Educational\\level}}
& High School or Less           & 11    & 5.0\%   \\
& Some College                  & 39    & 17.6\%   \\
& Undergraduate (BSc,~BA)       & 98    & 44.1\%   \\
& Graduate (MSc, PhD)           & 74    & 33.3\%   \\
\midrule
\multirow{6}{*}{\makecell[l]{Experience with\\OSS projects}}
& $<$1 year        & 31    & 13.9\%   \\
& 1-2 years        & 58    & 26.1\%   \\ 
& 2-3 years        & 49    & 22.1\%   \\
& 3-4 years        & 19    & 8.6\%   \\
& 4-5 years        & 12    & 5.4\%   \\
& $>$5 years       & 53    & 23.9\%   \\               
\midrule                
\multirow{6}{*}{\makecell[l]{Experience with\\closed-source\\ projects}}
& $<$1 year        & 53    & 23.9\%   \\ 
& 1-2 years        & 38    & 17.1\%   \\ 
& 2-3 years        & 34    & 15.3\%   \\
& 3-4 years        & 20    & 9.0\%   \\
& 4-5 years        & 8     & 3.6\%   \\
& $>$5 years       & 69    & 31.1\%   \\ 
\bottomrule
\end{tabularx}
\vspace{-2ex}
\end{table}

\subsubsection{SSATD Motivations}

\begin{figure*}[t]
\includegraphics[width=0.77\linewidth]{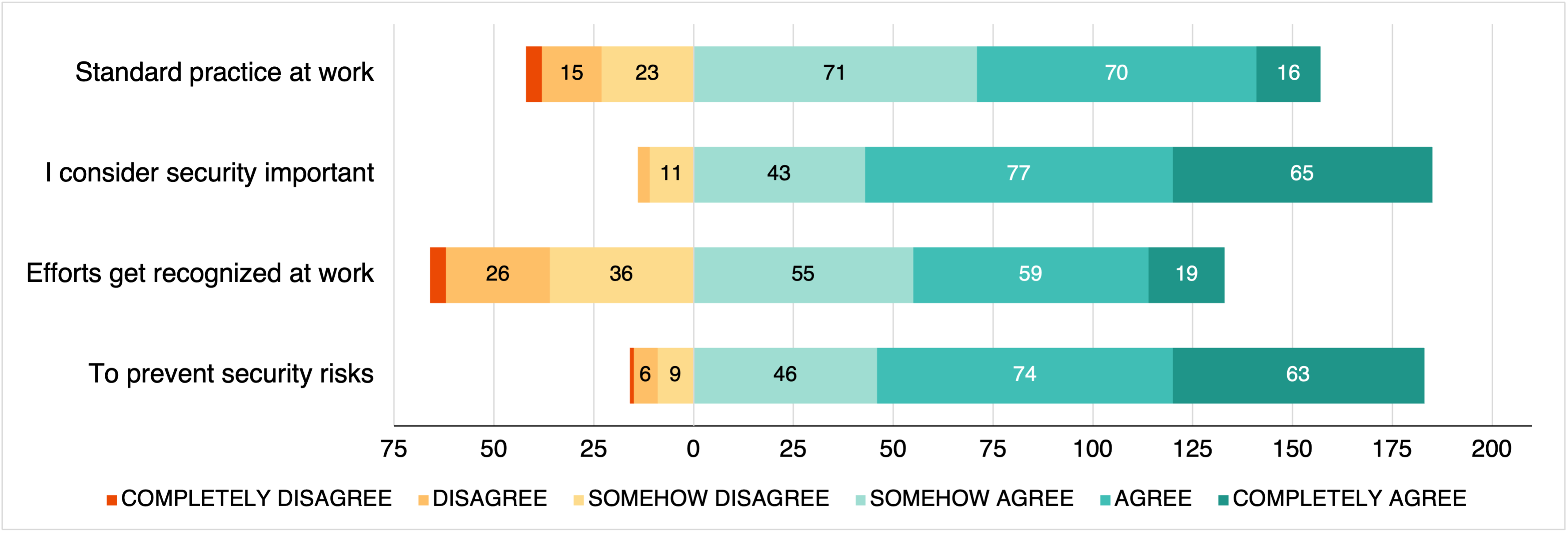}
\caption{Motivations to disclose security pointers in SATD. \underline{Note}: Frequencies < 10 are not labelled in the graph.}
\label{fig:motivations}
\end{figure*}

For the analysis of developers' motivations to disclose security pointers, we only considered the answers of those who claimed to have done so \textit{themselves} at some point (i.e., N=199 participants). As illustrated in Figure~\ref{fig:motivations}, most of them follow this practice because they either (i) consider security important or (ii) believe this can prevent potential security risks. Many also do so because it is standard practice at their workplace and, to a lesser extent, because such efforts get somehow recognised. However, these two factors, namely work practices and recognition, also gathered the highest number of disagreements among participants. Particularly, several developers tend to believe that placing security pointers inside code comments, commit messages, pull requests, or issue sections is not sufficiently recognised in their workplace.

We identified some additional motivations by analysing participants' answers to the open-ended question. For this, we followed an inductive method in which the first author performed open coding and, together with another author, discussed emerging themes and patterns in the data. In the case of any discrepancies, these were solved following a negotiated agreement approach \cite{campbell2013coding}. We found \textbf{5 extra motivations}, which we describe below and report their frequencies using the ($\times$) symbol.

\vspace{1ex}
\textit{\textbf{(i) Improve project quality} ($\times$17)} Many respondents stressed the importance of security pointers for the overall quality of OSS projects. For instance, they believed this was important to improve the performance of code review activities and the early detection of vulnerabilities.

\begin{quote}
    \textit{P60: ``Security pointers can help to improve the quality of code reviews by providing reviewers with information about potential security risks in the code being reviewed.''}
\end{quote}

\vspace{1ex}
\textit{\textbf{(ii) Comply with regulations and standards} ($\times$17) } Compliance with current standards and regulations also motivates developers to include security cues. Particularly, to demonstrate the adherence to legal requirements and security best practices.

\begin{quote}
    \textit{P212: ``The company I work for does not truly care about the benefits of good security pointers, instead the company is forced to do them due to compliance and regulation...''}
\end{quote}

\vspace{1ex}
\textit{\textbf{(iii) Facilitate collaboration} ($\times$51) } Many reported that security pointers help others to spot flaky code sections and fix them timely. Likewise, they believed these are important documentation instruments that ultimately promote and enhance collaboration among developers.

\begin{quote}
    \textit{P92: ``I use annotations and comments related to security in software as a note to collaborators and myself to be aware of these issues, especially if problems are easily introduced or something is confusing without more intimate knowledge of why something was done. That is, I add this information as form of documentation and to be proactive.''}
\end{quote}

\vspace{1ex}
\textit{\textbf{(iv) Self-reminders} ($\times$10) } Participants also mentioned that this helps them to keep track of the progress and status of their coding activities. Moreover, some also said it helps them to strengthen their secure development skills.

\begin{quote}
    \textit{P98: ``Just to remember what could be wrong in the future. I mean, there is a high probability that if I do not write something I will forget it in some months, when I may retake/reanalyse the project.''}
\end{quote}

\vspace{1ex}
\textit{\textbf{(v) Promote a security culture} ($\times$40) } Fostering a security-conscious culture across development teams was also a recurrent topic. That is, in terms of educating others and encouraging them to address software security in a proactive manner.

\begin{quote}
    \textit{P59: ``Educating my fellow contributors in producing better code, preventing mistakes if they modify my contributions and improving the community I am in by nudging them towards best practices.''}
\end{quote}

\subsubsection{SSATD Risks}

\begin{figure}[t]
\includegraphics[width=\linewidth]{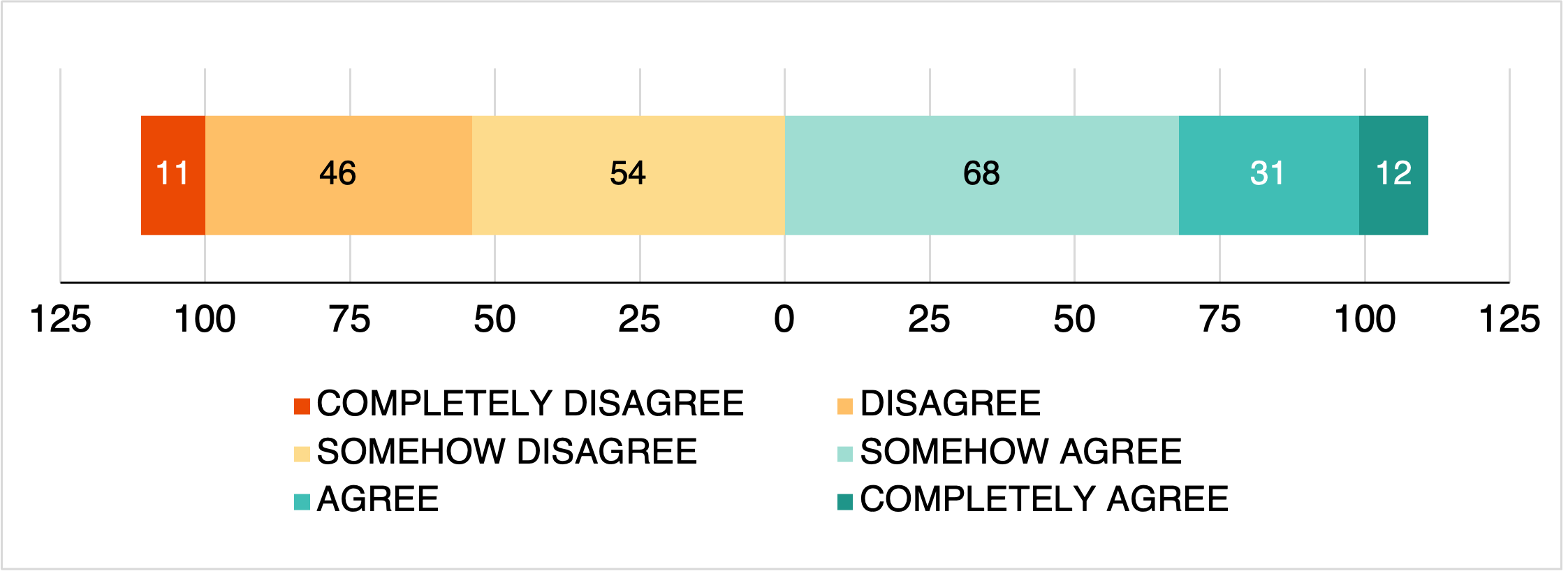}
\caption{Risk perceptions of security pointers in SATD.}
\label{fig:risk}
\end{figure}

Finally, we considered the answers of all participants when analysing their risk perceptions (i.e., N=222 answers). Figure~\ref{fig:risk} shows an even distribution of results, with 50\% of respondents tending to agree that security self-admissions may introduce risks in OSS projects. Conversely, the other 50\% reported having an opposite view in this regard. As with the motivations, we analysed the open-ended question about the potential risks of security pointers, following the same inductive approach to identify emerging topics. Overall, we gathered \textbf{3 different risks} that participants considered likely to emerge from the disclosure of security pointers.

\vspace{1ex}
\textit{\textbf{(i) Exposing vulnerabilities} ($\times$131) } A large number of respondents believed that security cues inside code comments, commit messages, pull requests, and issue sections may facilitate vulnerability exploits. Particularly, they considered that such pointers could create a roadmap of the security flaws inside OSS repositories and that attackers may easily take advantage of them.

\begin{quote}
    \textit{P60: ``Security pointers can make OSS more attractive to attackers, as they can provide a roadmap for exploiting vulnerabilities.''}\vspace{1ex}
    
    \textit{P91: ``If they are picked up by bad actors and exploited, this is risky.''}
\end{quote}

\vspace{1ex}
\textit{\textbf{(ii) Security misconceptions} ($\times$38) } Some participants said that pointers may not be accurate and lead either to a false sense of security or to deviate the attention of developers to non-existing vulnerabilities. Moreover, they considered that this sort of ``security advice'' could slow down the development process if not handled properly.

\begin{quote}
    \textit{P190: ``...if these tips are not clear or if they give the wrong advice, people might make mistakes and accidentally make the software less secure. So, it's really important that the tips are correct and easy to follow for them to be helpful.''}
\end{quote}

\vspace{1ex}
\textit{\textbf{(iii) Exposure of sensitive information} ($\times$21) } Another emerging topic concerned the possibility of revealing sensitive information to untrusted audiences, such as passwords, secrets, and credentials that unauthorized parties may exploit.

\begin{quote}
    \textit{P127: ``Security pointers may contain sensitive information such as passwords, API keys, or other credentials that can be exposed to unauthorized parties if not handled properly''}
\end{quote}

\section{Discussion}\label{sec:discussion}

In this section, we discuss our findings in the context of existing literature and provide implications for research and practice.


\subsubsection*{\textbf{SSATD and Software Security.}} The outcome of our study provides valuable insights into the role and impact of SSATD on software security. On the one hand, we observe that security pointers disclosed in the form of commit messages, pull requests, issue sections, and code comments can provide enough detail to indicate the presence of vulnerabilities. Furthermore, as reported in Section~\ref{sec:res_pointers}, the level of granularity of many of these self-admissions can suggest the existence of dangerous software weaknesses (i.e., CWEs listed in MITRE's Top-25). In this sense, our results align with the ones of \citet{russo2022msr}, who identified 14 different CWE issues associated with SATD instances found in the Chromium C-Code project (i.e., in the form of code comments). Moreover, a recent work by \citet{pan2022fse} revealed that issue reports gathered across 1.390 OSS projects from GitHub contained information about 132 different CWE vulnerability types. \cready{Regarding its distribution across TD types, a study by \citet{bavota2016satd} concludes that most SATD corresponds to code/design debt, followed by defect and requirement debt. We observe a similar pattern in the SSATD dataset where code/design debt accounts for 58.6\% of the code comments, defect for 24.15\%, and requirement for 17.25\%.}

Also in line with prior findings, our results account for the threats that security pointers may pose for the OSS ecosystem if disclosed early to the public. That is, before vulnerabilities get reported privately to the corresponding project maintainers. Particularly, malicious actors could take advantage of the leaked information (namely, the reproducible steps of a vulnerability) and launch zero-day attacks \cite{pan2022fse}. Furthermore, such a practice can severely affect commercial closed-source software solutions, as they often use OSS resources (e.g., packages and libraries) in their deployment~\cite{ladisa2023sok}. Hence, current vulnerability disclosure protocols (e.g., ISO/IEC 29147:2018 \cite{ISO29147}) should acknowledge the side effects of SSATD to become more resilient against this type of attack.


\subsubsection*{\textbf{SSATD and Privacy Behaviour}} The insights gathered through the online survey are also nuanced. We can see that most developers are keen to report security issues throughout different SATD sources but, at the same time, seem to acknowledge that this is a risky practice (especially when working in an OSS environment). Prior work by \citet{diaz2023chase} suggests that software practitioners tend to underestimate the potential negative consequences of revealing personal and project-related information in social coding platforms and may exhibit paradoxical behaviour in this regard. That is, they often engage in unsavvy privacy/security practices despite their concerns. In part, such behaviour can be associated with their perceptions of trust in the OSS ecosystem and prior experiences of privacy invasion \cite{diaz2023chase}. In this sense, those who have not suffered from any privacy harm or consider OSS platforms as safe collaboration channels are prone to reveal more information about themselves or the projects they work on. Previous findings concerning software documentation \cite{shmerlin2015document,vidoni2021self} and security best practices \cite{assal2019think,rauf2022influences} show that extrinsic/intrinsic motivations (e.g., self-reminders, peer pressure and organisational culture) also play an important role in this process. Overall, our results align with the ones reported in the current literature and show that such motivations can also foster SSATD. Therefore, it is important to balance out these drivers with higher levels of risk awareness to promote the incorporation of security pointers within certain confidentiality boundaries.





\begin{figure}[b]
\begin{GrayBox}
\textbf{RESEARCH OUTLOOK.}
\begin{enumerate}[leftmargin=3ex]
\item Security pointers disclosed inside SATD sources can provide enough detail to suggest the presence of well-known software vulnerabilities (i.e., listed as CWEs).\vspace{1ex}
\item Vulnerability disclosure protocols should take SSATD into account to better safeguard both commercial and open-source software projects against zero-day attacks. \vspace{1ex}
\item Developers should gain awareness of the risks of disseminating security pointers inside SATD sources and count on means for reporting them safely to trusted parties only.\vspace{1ex}
\item ML solutions for automatically predicting software vulnerabilities should consider improving their performance by incorporating SSATD sources as model features.\vspace{1ex}
\item \cready{Approaches to TD prioritisation and repayment should take SSATD explicitly into consideration when looking for security weaknesses with potential TD impact.}
\end{enumerate}
\end{GrayBox}
\end{figure}

\subsubsection*{\textbf{Sensitive Information in SSATD}} From the qualitative analysis of the open-ended questions, we observe concerns stemming from the unintended disclosure of sensitive/exploitable information to untrusted audiences. These concerns are closely related to \textit{secret sprawling}, an emerging phenomenon in software engineering where secrets (e.g., passwords, certificates, and authentication tokens) get accidentally pushed/committed into publicly-available OSS repositories \cite{krause2023pushed}. Particularly, it is estimated that secret sprawling on GitHub has more than doubled since 2020, making it a significant security threat for software projects \cite{gguardian2022}. In turn, several commercial and OSS tools have been proposed to facilitate the ``sanitization'' of code repositories through the automatic identification of secrets \cite{lounici2021optimizing}. Still, our research shows that this is a multifaceted, complex problem, and security self-admissions can also enlarge a system's attack surface. This calls for developer-centred tools and instruments that could help prevent the dissemination of SSATD to untrusted parties by preserving the contextual integrity of security pointers. 

\subsubsection*{\textbf{SSATD in Vulnerability Prediction Methods}} As mentioned in Section~\ref{sec:background}, recent work has started investigating whether TD indicators can improve the performance of ML-based vulnerability prediction methods \cite{siavvas2022technical}. Our findings also contribute in this regard as they could be leveraged to enhance these approaches. That is, by using SSATD as a complementary feature in ML models for automatically detecting security weaknesses. Moreover, as we reported in Section~\ref{sec:res_pointers}, issue sections have been shown to concentrate the vast majority of the security pointers in our dataset. Therefore, future investigations in ML for security should strongly consider expanding the scope of their analysis to multiple SSATD sources and explore their suitability for developing more accurate automatic vulnerability detection solutions. Still, further analysis is required to check whether these pointers effectively refer to vulnerable code implementations and account for a significant number of true positive cases.

\subsubsection*{\cready{\textbf{TD Prioritisation and Repayment}}} \cready{One of the challenges of TD management is related to its prioritisation and later repayment. Given the high repayment cost of security vulnerabilities, recent work has started to explore efficient ways to identify, assess and manage security-related TD in software projects. These include, for example,  risk-based management frameworks \cite{rindell2019managing} and quality models to operationalise the scoring of TD instances associated with CWE violations \cite{izurieta2018td}. Nevertheless, such approaches often concentrate on TD items that can be spotted via security engineering techniques and tools (e.g., bad coding practices or structural deficiencies), leaving SATD out of scope. The insights gained in this work show that security information contained in SATD can help characterise the attack surface of software projects. Hence, they could enhance current approaches for TD prioritisation and repayment like the one proposed by \citet{izurieta2018td}, which take CWE violations explicitly into account when computing 
a relative ranking of security weaknesses with a potential TD impact.} 








\section{Threats to Validity}\label{sec:validity}

To a certain extent, the results of our study are subject to limitations related to its experimental design. In particular, the following \textit{construct}, \textit{external}, and \textit{internal} threats may affect the validity of our findings and conclusions:

\textbf{Internal Validity}: As described in Section~\ref{sec:methodology}, we conducted a keyword-based search to identify SSATD candidates in the reference dataset. Such an approach is not error-free and may produce some false-positive results. For instance, security keywords like `hack' and `hash' triggered many cases in which the candidates did not contain any security pointer (e.g., ``\texttt{Hack to allow external control}''). Likewise, since the keyword list we employed is not exhaustive, some true positive cases may have been overlooked. We sought to minimize these threats by performing a manual check of each SSATD candidate and employing an extensive list of security keywords. As mentioned earlier, the check was conducted by two of the paper co-authors, who are both knowledgeable in cybersecurity and solved their discrepancies through negotiation. As for the keywords, we used a list that is well-established and documented in the current literature. Still, we acknowledge we may have missed some SSATD instances throughout the data curation process.

The recruitment process followed in the survey study also suffers from limitations. Particularly, it is known that prospective participants recruited via crowd-sourcing platforms may not always meet the study eligibility criteria (e.g., some of them may claim to have prior OSS development experience when, in fact, they do not). We have addressed this issue by (i) using Prolific's in-built qualification features and (ii) running a screening questionnaire before the main survey, as suggested when conducting studies of this nature (see \cite{tahaei2022recruiting}). \cready{Additionally, we introduced attention questions to spot random answers from dishonest participants. However, we did not check for participants' degree of cybersecurity expertise as it was not our intention to capture the perspective of security experts only (see RQ3). Still, some basic knowledge in this regard may have been beneficial to elicit more relevant responses}.

Overall, the detailed answers observed in the open-ended questions gave us the confidence that the vast majority of participants had the right level of expertise. Still, this recruitment approach may also introduce a ``survivorship bias'' in the studied sample, failing to collect the views from those OSS practitioners outside Prolific. This threat was mitigated thanks to the answers of several experienced participants within our sample (Section~\ref{sec:survey_results}).


\textbf{Construct Validity}: The manual mapping of CWE identifiers to SSATD instances is prone to subjective judgements and may lead to spurious conclusions. Unlike the labelling of SSATD instances, this process is more complex, given the high number of CWE identifiers available. Therefore, we followed an opportunistic approach in which two authors browsed the CWE database guided by the keywords found on each SSATD item (Section~\ref{sec:mthd_val}). This process was first performed by one of them and then validated by the other one. Follow-up discussions between the two sought to enhance the construct validity at this step. For dubious cases, a third author (also with extensive cybersecurity expertise) was consulted to facilitate reaching a decision and further improve the labelling outcome. We followed a similar approach to mitigate subjective biases while coding the survey's open-ended questions (i.e., Q7 and Q9). Here, one author performed an initial coding of the data, whereas another checked them and completed/fixed them when needed.


\textbf{External Validity}: Although the reference dataset used in the repositories study is fairly large (94,455 software artefacts), it may not offer a complete picture of the SATD phenomenon in OSS. Moreover, our SSATD dataset contains 201 instances, which may pose a potential threat to the generalizability of our results. We have addressed this concern by comparing and discussing our findings with the ones reported in the current literature to minimize the risk of interpretation bias. Generalizability threats also arise from the survey study, given the size and composition of the studied sample. On the one hand, because of its size, we cannot generalise the study results to the entire community of OSS developers. Instead, the insights gained from it should be seen as preliminary and motivate further research in this area. On the other hand, because of its demographic composition, we may have failed to capture the perspectives of underrepresented gender groups (e.g., women and gender-diverse people). Hence, future investigations should target larger and more gender-diverse samples for the sake of generalizability.


\section{Conclusion and Future Work}\label{sec:conclusion}

The findings gathered throughout our mixed-methods study suggest that security pointers disclosed across SATD sources are like a double-edged sword. On the one hand, they aid developers in finding and fixing security weaknesses spread across OSS projects. Furthermore, they are often considered important for fostering a security culture across development teams and complying with current standards and regulations. However, we also see that practitioners often engage in this practice despite the potential risks, particularly concerning the role of SSATD in identifying exploitable vulnerabilities.

Having mapped several SSATD instances to concrete CWEs speaks about the severity of spreading security pointers carelessly across OSS repositories. Hence, preserving the contextual integrity of this information is critical to protect both commercial and OSS solutions against zero-day attacks. As discussed in Section~\ref{sec:discussion}, vulnerability disclosure protocols should take SSATD explicitly into account to enhance their robustness against this type of attack. Likewise, secret scanning tools should extend their scope with SSATD identification features to assist developers in keeping this information only accessible to trusted parties. Last but not least, developers' awareness and training about the inherent risks of this practice are key for promoting best practices in this regard.

Several directions for future work arise from the results of this study. One relates to the role of privacy-related antecedents when disclosing security-related information across SATD artefacts. Particularly, prior work has reported that developers' perceptions of trust, risk, and control over personal and project-related information may influence their privacy practices inside Social Coding Platforms (SCPs) like GitHub \cite{diaz2023chase}. Since SCPs are major sources of OSS repositories, it would be relevant to determine whether such perceptions contribute (or not) to the dissemination of SSATD across code comments, commit messages, issue sections and pull requests. In addition, a closer inspection of the code sections linked to SSATD instances would help us to get a more complete picture of their actual implications. That is, whether they are good indicators of vulnerabilities and we can leverage them to elaborate ML models for their automatic prediction.

\cready{Another direction for future research concerns the application of advanced Natural Language Processing (NLP) techniques to support the automatic identification of security-relevant SATD instances. Prior work has elaborated on the use of ML for detecting and classifying SATD into multiple TD types, such as requirement, code, documentation and design \cite{sharma2022self, li2023automatic, santos2020long}. Some of the ML methods that have been applied successfully in the past include Long Short-Term Memory Neural Networks (LSTM) \cite{santos2020long}, Convolutional Neural Networks (CNN) \cite{li2023automatic}, and  Pre-Trained Language Models such as RoBERTa \cite{sharma2022self}. These techniques could help overcome, to a great extent, the limitations of a keyword-based search discussed in Section~\ref{sec:validity}. Moreover, many of these models, which target generic SATD classification, could be easily customised using transfer learning to identify SSATD automatically.}



\section*{Replication Package}\label{sec:replication}

Survey instruments, consent forms, and study results are available at \url{https://zenodo.org/doi/10.5281/zenodo.10551751}

\begin{appendices}

\section{Survey Questions} \label{appendix}

The most salient questions from the survey described in Section~\ref{sec:survey_study} are listed below. Note that open-ended questions are indicated with an asterisk (*). The complete survey instrument is available in the paper's \hyperref[sec:replication]{replication package}.\vspace{2mm}

\begin{enumerate}[leftmargin=*]
    \item \textbf{SSATD Prevalence:}
    \begin{itemize}[leftmargin=*]
    \item \textbf{Q1.} How often have you (yourself) produced an artefact (i.e., commits, issues, comments, or pull requests) containing security pointers?
    \item \textbf{Q2.} How often have you seen others producing an artefact containing security pointers?
    \end{itemize}\vspace{1mm}
    \underline{Answer options}: \textit{Never, Rarely, Sometimes, Often, Always}.\vspace{3mm}

    \item \textbf{SSATD Motivations:}
    \begin{itemize}[leftmargin=*]
    \item \textbf{Q3.} While working in OSS projects, I introduce security pointers inside an artefact because this is a standard practice at work.
    \item \textbf{Q4.} While working in OSS projects, I introduce security pointers inside an artefact because I consider security an important aspect.
    \item \textbf{Q5.} While working in OSS projects, I introduce security pointers inside an artefact because such efforts are recognized at work.
    \item \textbf{Q6.} While working in OSS projects, I introduce security pointers inside an artefact because they help prevent security risks.
    \item \textbf{Q7*.} What other motives may drive you to introduce security pointers inside an artefact?
    \end{itemize}\vspace{1mm}
    \underline{Answer options}: \textit{Completely Disagree, Disagree, Somehow Disagree, Somehow Agree, Agree, Completely Agree}.\vspace{3mm}

    \item \textbf{SSATD Risks:} 
    \begin{itemize}[leftmargin=*]
    \item \textbf{Q8.} I think security pointers inside an artefact may introduce risks in the OSS being deployed.
    \item \textbf{Q9*.} What kind of risks do you think security pointers inside an artefact introduce in the OSS being deployed?
    \end{itemize}\vspace{1mm}
    \underline{Answer options}: \textit{Completely Disagree, Disagree, Somehow Disagree, Somehow Agree, Agree, Completely Agree}.
\end{enumerate}

\end{appendices}


\section*{Acknowledgements}

This research has received funding from the European Union's Horizon 2020 research and innovation programme, within the OpenInnoTrain project under the Marie Sk\l{}owdowska-Curie grant agreement no. 823971. The content of this publication does not reflect the official opinion of the European Union. Responsibility for the information and views expressed in the publication lies entirely with the author(s).
\bibliographystyle{ACM-Reference-Format}
\bibliography{references}

\end{document}